\documentclass[prstab,letterpaper,showpacs,final,twocolumn]{revtex4-1}
\usepackage{psfrag}
\usepackage{graphicx}
\usepackage{dcolumn}
\usepackage{amsmath}

\begin{document}

\title{Active Negative Index Metamaterial Powered by an Electron
Beam}
\author{M.~A.~Shapiro$^1$, S.~Trendafilov$^2$, Y.~Urzhumov$^3$, A.~Alu$^4$,
R.~J.~Temkin$^1$, and
G.~Shvets$^2$}\email{gena@physics.utexas.edu}
\affiliation{$^1$Plasma Science and Fusion Center, Massachusetts
Institute of Technology, Cambridge, MA 02139\\$^2$Department of
Physics, The University of Texas at Austin, Austin TX
78712\\$^3$Center for Metamaterials and Integrated Plasmonics,
Pratt School of Engineering, Duke University, Durham, NC
27708\\$^4$Department of Electrical and Computer Engineering, The
University of Texas at Austin, Austin TX 78712}

\date{\today}

\begin{abstract}
A novel active negative index metamaterial that derives its gain
from an electron beam is introduced. The metamaterial consists of a
stack of equidistant parallel metal plates perforated by a
periodic array of holes shaped as complementary split-ring
resonators. It is shown that this structure supports a
negative-index transverse magnetic electromagnetic mode that can
resonantly interact with a relativistic electron beam. Such
metamaterial can be used as a coherent radiation source or a
particle accelerator.
\end{abstract}

\pacs{81.05.Xj, 41.60.Bq, 41.75.Lx, 07.57.Hm}

\maketitle

\begin{figure}[hbt]
\centering
\includegraphics[width=0.23\textwidth]{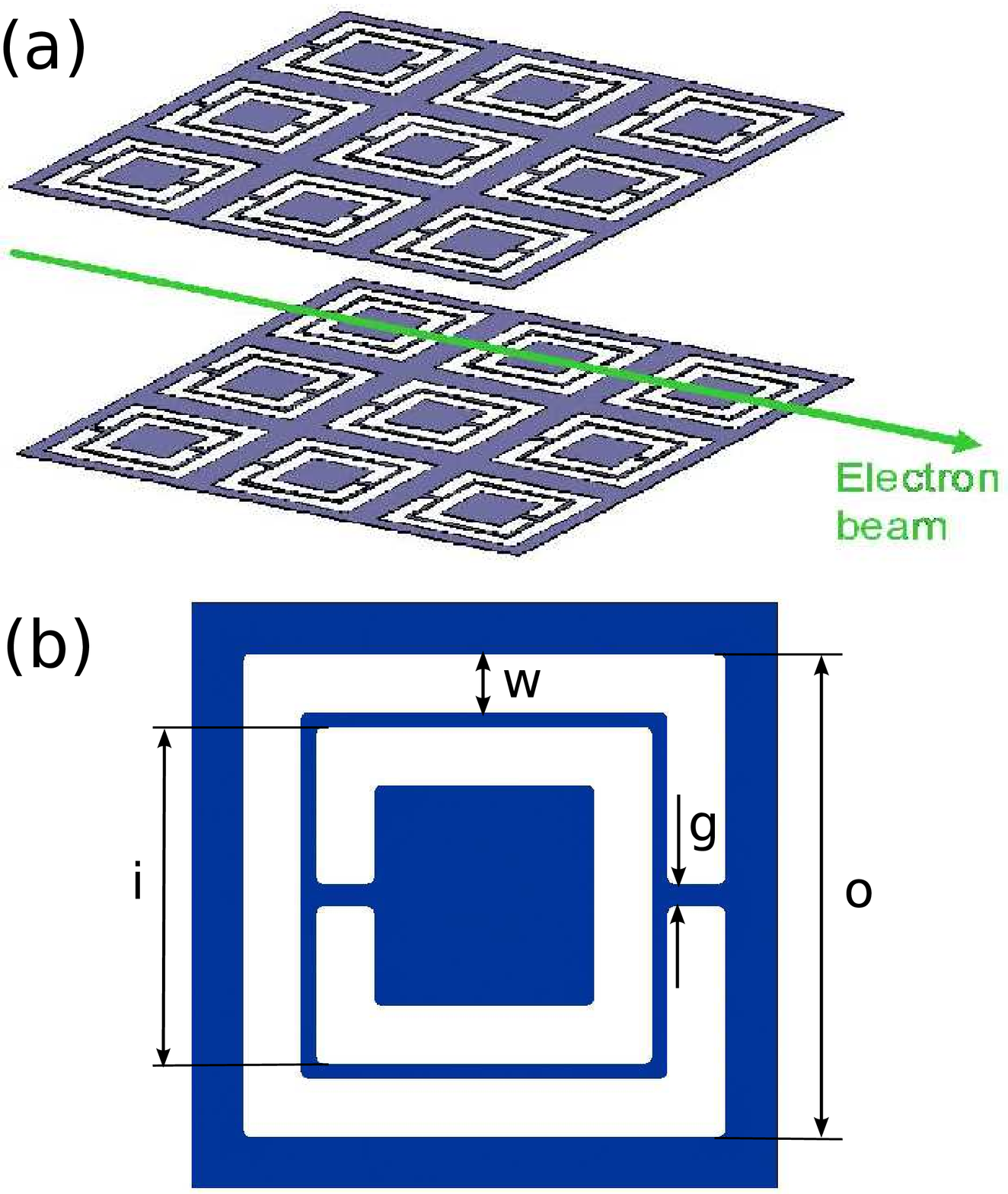} \
\includegraphics[width=0.23\textwidth]{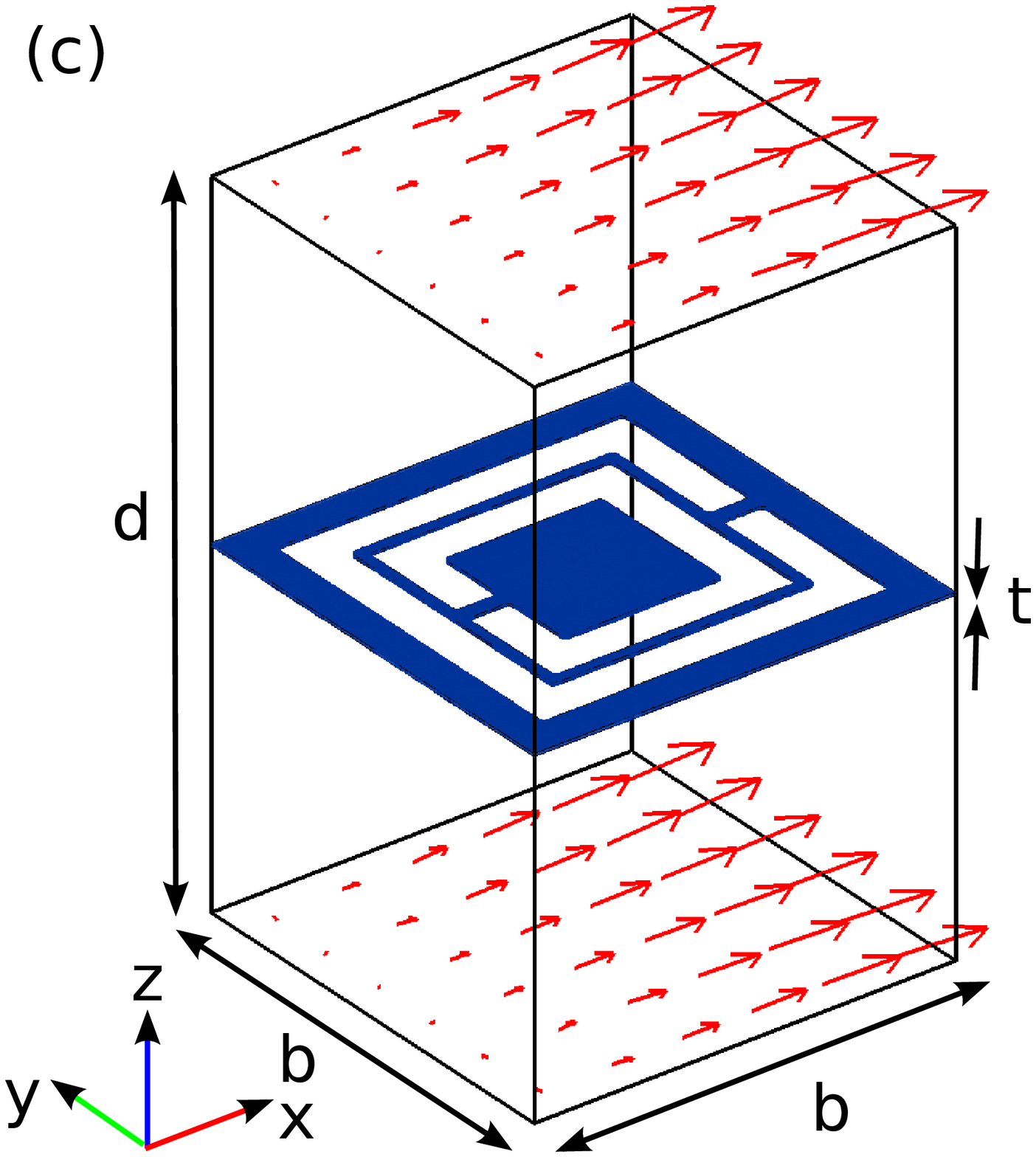}
\caption{(Color online) (a) Schematic of a Negative Index
Meta-Waveguide (NIMW) comprised of a stack of metal plates
patterned by complementary split-ring resonators (C-SRRs). An
electron beam propagating in the $x$-direction and interacting
with the NIMW is also shown. (b) C-SRR's dimensions: outer ring
slot length $o$=6.6$mm$; inner ring slot length $i$=4.6$mm$; slot
width $w$=0.8$mm$; gap width $g$=0.3$mm$. (c) Single cell of a
NIMW and midplane electric fields (arrows) interacting with the beam:
stacking distance between metal planes $d$=12.8 $mm$; square
lattice period $b$=8 $mm$; metal thickness $t$=0.05$mm$. These
dimensions were chosen for a frequency near $f_0=5$GHz as
described in detail below.}\label{fig:schematic}
\end{figure}

Artificially structured metamaterials (MTMs) possess exotic macroscopic
electromagnetic properties that cannot be achieved in natural
materials. Constructed from simple planar elements such as
split-ring resonators and thin wires~\cite{smith_prl00}, MTMs
enable a variety of applications such as ``perfect'' lenses,
compact transmission lines and antennas, electromagnetic cloaks,
and many others
~\cite{pendrylens_prl00,caloz_ieee05,ziolkowski_06,smith_cloak_06}.
Negative refractive
index~\cite{veselago_68,smith_prl00,zhang_nature08,atwater_science07}
is one of the most surprising and thoroughly studied properties
enabled by MTMs. In this Letter we describe a new class of
negative index MTMs that can strongly interact with an electron
beam, thereby opening new opportunities for vacuum electronics
devices such as coherent radiation sources and particle
accelerators. The specific implementation of such a negative-index
meta-waveguide (NIMW) analyzed in this Letter and schematically
shown in Fig.~\ref{fig:schematic} is obtained by patterning an
array of split-ring resonator cutouts on the plates of a stack of
planar metallic waveguides.

The NIMW belongs to the category of complementary metamaterials
(C-MTMs) \cite{falcone}. C-MTMs utilize the complements of the
traditional split-ring resonators (SRR) in order to achieve a
complementary electromagnetic response: an SRR exhibits a strong
magnetic response while a C-SRR has a strong electric response.
Narrow waveguides patterned with C-SRRs have been
used~\cite{smith_prl08} to demonstrate enhanced tunneling of
transverse electromagnetic (TEM-like) waves. In this Letter we
demonstrate that this structure supports a negative-index
transverse magnetic (TM) mode: an electromagnetic mode propagating
in the $x$-direction, with $E_x$ being the only non-vanishing
component in the waveguide's mid-plane at $z=0$. As demonstrated
below, the negative effective permittivity of the NIMW
$\epsilon_{\rm eff}<0$ is imparted to it by resonant
C-SRRs~\cite{falcone,smith_prl08}, while the negative effective
permeability $\mu_{\rm eff}<0$ is due to the transverse
confinement of the TM modes~\cite{shvets_prb03} supported by the narrow
(width $d$ in the $z$-direction is much smaller than the
wavelength $\lambda \equiv 2\pi c/\omega$) waveguides formed by the neighboring plates.
The importance of utilizing TM modes lies in their ability to
resonantly interact via finite $E_x$ with relativistic electron
beams when their phase velocity $v_{ph} \equiv \omega/k_x$ is
equal to the beam's velocity $v_b$. Such interaction can be
exploited to either transfer the electromagnetic energy to the
beam (particle accelerator) or to extract energy from the beam
(coherent radiation source).

The attraction of the NIMW for coherent high-frequency radiation
generation is four-fold. First, the opposite sign of the group
velocity and the beam velocity can result in an instability
utilized in backward-wave oscillators (BWO) or (for lower beam
currents) amplifiers (BWA)~\cite{tsimring}. The sub-wavelength
nature (lateral period $b \ll \lambda$) of the NIMW supported by
its resonant C-SRRs distinguishes it from the traditional BWOs
which rely on the interaction between an electron beam and a
spatial harmonic of the electromagnetic field in a periodic
structure. Second, the low group velocity $v_g << c$ of the
negative-index waves due to the NIMW's resonant C-SRRs increases
spatial gain, reduces the starting current requirement of a BWO,
and enables shorter structures. Third, NIMW's constitutive
elements (C-SRRs) can be produced using standard planar
fabrication techniques. This is particularly advantageous for the
generation of THz and millimeter waves because the fabrication of
conventional BWOs~\cite{goebel_ieee94} relies on high-precision
machining that becomes challenging for shorter wavelengths.
Finally, the output radiation frequency of a NIMW can be
accurately and rapidly controlled by electric or optical tuning of
the resonant frequency of the
C-SRRs~\cite{padilla_nature06}. We note that the
absolute instability of electron beams inside a negative-index
medium has been suggested earlier~\cite{bliokh}, albeit limited to
a hypothetical isotropic negative index material.

\begin{figure}[hbt]
\centering
\includegraphics[width=65mm]{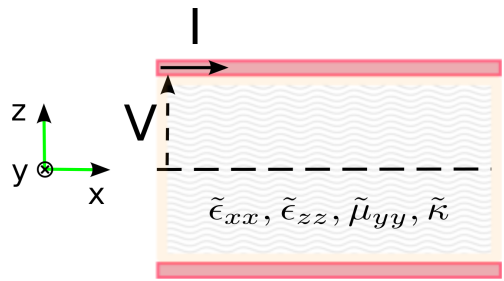}
\caption{(Color online) Side view of a parallel plate waveguide
with effective filler medium. Voltage $V$ and current $I$ are used
for extracting the MTM's constitutive parameters.}
\label{fig:waveguide}
\end{figure}

Below we demonstrate that the NIMW shown in
Fig.~\ref{fig:schematic} can be properly described as an effective
bianisotropic~\cite{marques_prb02} negative index medium for
electromagnetic waves propagating in the $x$-direction. By
restricting the macroscopic (i.e. properly averaged over the
metamaterial's unit cell) electromagnetic field components to
$\bar{E}_z$ and $\bar{H}_y$, such metamaterial can be
characterized by a set of constitutive parameters $\epsilon_{\rm
eff}$, $\mu_{\rm eff}$, and the bianisotropy coefficient
$\kappa_{\rm eff}$ defined according to
\begin{equation}\label{eq:constitutive}
    \bar{D}_z = \epsilon_{\rm eff} \bar{E}_z - i \kappa_{\rm eff}
    \bar{H}_y, \ \ \ \bar{B}_y = + i\kappa_{\rm eff} \bar{E}_z +
    \mu_{\rm eff} \bar{H}_y,
\end{equation}
and can be shown~\cite{marques_prb02,ozbay} to support
electromagnetic waves propagating with refractive index $n$ given
by
\begin{equation}\label{dispersion}
    n \equiv \frac{c k_x}{\omega} = \pm
    \sqrt{\epsilon_{eff}\mu_{eff}-\kappa_{eff}^2},
\end{equation}
where the negative sign is assigned to the propagating waves with
$\epsilon_{\rm eff},\mu_{\rm eff} < 0$. To understand the
emergence of negative-index waves in a NIMW, we first examine the
origin of $\mu_{\rm eff} < 0$ for wave propagation through a
metamaterial composed of an array of planar waveguides with
perfectly electrically conducting (PEC) walls stacked along the
$z$-direction. It is well established that a metamaterial composed by an array of 
metallic plates may be rigorously homogenized by studying the modes excited between 
any two neighboring plates~\cite{rotman,silveirinha}. Here we apply this homogenization 
model to the parallel-plate metamaterial of Fig.~\ref{fig:schematic}a. 
A single cell of the array is shown in Fig.~\ref{fig:waveguide}. 
The effective metamaterial properties are determined by the dominant TM mode, with
wavenumber $k_x$ and frequency $\omega$. We note that the longitudinal component of 
the electric field, enabled by the parallel-plates, can resonantly interact with an
electron beam propagating in the $x$-direction. The TM$_1$ mode is
symmetric with respect to the $z=0$ mid-plane, and possesses
non-vanishing fields $E_x$ (even function of $z$), $E_z$, and
$H_y$ (both odd functions of $z$).

In anticipation of the need to emulate the effects of C-SRRs and
the electron beam, the waveguide is assumed to be filled with a
material characterized by a permittivity tensor with
non-trivial components $\tilde{\epsilon}_{xx}$ and
$\tilde{\epsilon}_{zz}$,
and a single non-vanishing component
$\tilde{\kappa}$ defined as in Eq.(\ref{eq:constitutive}).
Finite $\tilde{\kappa}$ emulates magneto-optical coupling
introduced by the C-SRR~\cite{marques_prb02},
$\tilde{\epsilon}_{zz} \neq 1$ emulates resonant electric response
of the C-SRR, while $\tilde{\epsilon}_{xx} \neq 1$ emulates the
wave's interaction with an electron beam when the resonance
condition $\omega = k_x v_b$ is satisfied.

The effective constitutive parameters may be computed by analyzing the 
propagation properties of the dominant TM$_1$ mode, using the
transmission-line characteristic impedances for forward and
backward propagating TM waves according to $Z_{ch}^{\pm}\equiv \pm
V/I$, where the transmission line's voltage $V$ and current $I$
are defined according to
\begin {eqnarray}
\label{I_and_V} I=-\int_{-b/2}^{b/2} dy
H_{y}(x=\mp b/2, z=t/2)\\
V=\frac{1}{b} \int_{-b/2}^{b/2} dy \int_{t/2}^{d/2} dz E_{z}(x=\mp
b/2), \nonumber
\end {eqnarray}
and the top and bottom signs correspond to the forward and backward
waves, respectively. While $Z_{ch}^{+} = Z_{ch}^{-}$ for an air filled
transmission line shown in Fig.~\ref{fig:waveguide}, that would no
longer be the case when magnetoelectric coupling is present in the
filling medium, as would be the case in the more general
bianisotropic structure shown in Fig.~\ref{fig:schematic}.
Effective material parameters can then be obtained from the
transmission-line model through
\begin{eqnarray}\label{eq:effective_params_TL}
\epsilon_{eff}&=& \frac{ck_x}{\omega}\frac{2}{Z_{ch}^{+} +
Z_{ch}^-} {Z_0}, \nonumber
\\ \mu_{eff}&=&\frac{ck_x}{\omega}\frac{2}{Y_{ch}^{+} +
Y_{ch}^{-} } \frac{1}{Z_0}, \nonumber \\ \kappa_{eff}&=&i
\frac{Y_{ch}^+ - Y_{ch}^-}{2} Z_0 \mu_{eff},
\end{eqnarray}
where $Z_0=377$Ohm is the free-space impedance, and
$Y_{ch}\equiv1/Z_{ch}$ is the characteristic admittance of the
transmission line.

Applying the above definitions of effective parameters and
characteristic impedances to the $TM_1$ mode of 
the conventional parallel-plate metamaterial in Fig.~\ref{fig:waveguide} 
made of smooth metallic plates and suitable filler medium, 
we obtain:
$\kappa_{eff} = \tilde{\kappa}$ and
\begin{equation}\label{eq:effective_params_WG}
    \epsilon_{eff} = \tilde{\epsilon}_{zz}\frac{\pi b}{d}, \  \mu_{eff} =
    \left(\tilde{\mu}_{yy} - \left( \frac{\pi/d}{\omega/c} \right)^2
    \frac{1}{\tilde{\epsilon}_{xx}}\right)\frac{d}{\pi b},
\end{equation}
resulting in the dispersion relation for the TM$_1$ wave:
\begin{equation}\label{eq:dispersion_WG}
    \frac{ck_x}{\omega} = \pm \sqrt{ \tilde{\epsilon}_{zz}
    \tilde{\mu}_{yy} - \tilde{\kappa}^2 -
    \left(\frac{\pi/d}{\omega/c}\right)^2
    \frac{\tilde{\epsilon}_{zz}}{\tilde{\epsilon}_{xx}}}.
\end{equation}

Several insights can be gained from
Eq.~(\ref{eq:effective_params_WG}). First, the effective magnetic
permeability turns negative for $\omega < \omega_c$, where
$\omega_c = c\pi/d$ is the cutoff frequency of the considered
TM$_1$ mode. Therefore, one approach to achieving negative-index
propagation at $\omega < \omega_c$ is to pattern the waveguide's
wall in such a way as to ensure that
$\tilde{\epsilon}_{zz}(\omega) < 0$. Second, if
$Im{(\tilde{\epsilon}_{xx})} \neq 0$ (as is the case for a beam
resonantly interacting with the $E_x$ component of the mode), then
$Im{(\mu_{\rm eff})} \neq 0$, resulting in an active (gain)
metamaterial. That the longitudinal component of the electric
field $E_{x}$ (and, therefore, $\tilde{\epsilon}_{xx}$)
contributes to the effective magnetic permeability $\mu_{\rm eff}$
of confined TM modes has been
known~\cite{shvets_prb03,atwater_science07} from theoretical and
experimental studies, but the possibility of employing an electron
beam for controlling the imaginary part of $\mu_{\rm eff}$ and
realizing gain in metamaterials has not been recognized. Finally,
Eq.~(\ref{eq:dispersion_WG}) can be recast in the conventional
form for the theory of travelling wave tubes
(TWTs)~\cite{tsimring} by assuming that the waveguide is filled
with an active medium with permittivity
$\tilde\epsilon_{xx}^{(b)}=1-\omega_{b}^2/(\omega-k_{x}v_{b})^2$,
where $\omega_{b}$ is the electron beam plasma frequency. The
resulting dispersion relation for the {\rm active} NIMW can now be
re-written as
\begin{eqnarray}\label{eq:disp_eq_beam}
    &&\left(k_{x}^{2} - \frac{\omega^{2}}{c^{2}}
    (\epsilon_{eff}\mu_{eff} - \kappa_{eff}^{2}) \right)
    \left( \omega-k_{x}v_{b} \right)^{2} = \nonumber \\
    &&\frac{\omega^2}{c^2}(\mu_{eff}-1)\epsilon_{eff}
    \omega_{b}^{2},
\end{eqnarray}
where, because of the wave-beam interaction, the frequency
$\omega$ is a complex number for a real propagation constant
$k_{x}$. Analogous to the linear theory of the
TWT~\cite{tsimring}, Eq.~(\ref{eq:disp_eq_beam}) is quartic in
$\omega$ having four complex roots that represent three forward
waves (with positive, negative, and zero gain) and one backward
wave (not affected by the beam). The maximum gain $\gamma_{\rm
max} = \sqrt{3}/2 \rho$ is achieved at the beam-wave synchronism
$\omega_{\rm NIMW}(k_x) = k_{x}v_{b}$ (zero detuning) condition,
where $\omega_{\rm NIMW}(k_x)$ is the dispersion relation without
the beam, and the Pierce parameter~\cite{tsimring} of the NIMW is
given by
\begin{equation}\label{eq:max_incr}
\rho = \left(\frac{1}{2} \frac{v_{b}^2}{c^2} (\mu_{eff}-1)
\epsilon_{eff} \omega_{\rm NIMW} \omega_{b}^{2}
\right)^{\frac{1}{3}}.
\end{equation}

\begin{figure}[hbt]
\centering
\includegraphics[width=0.45\textwidth]{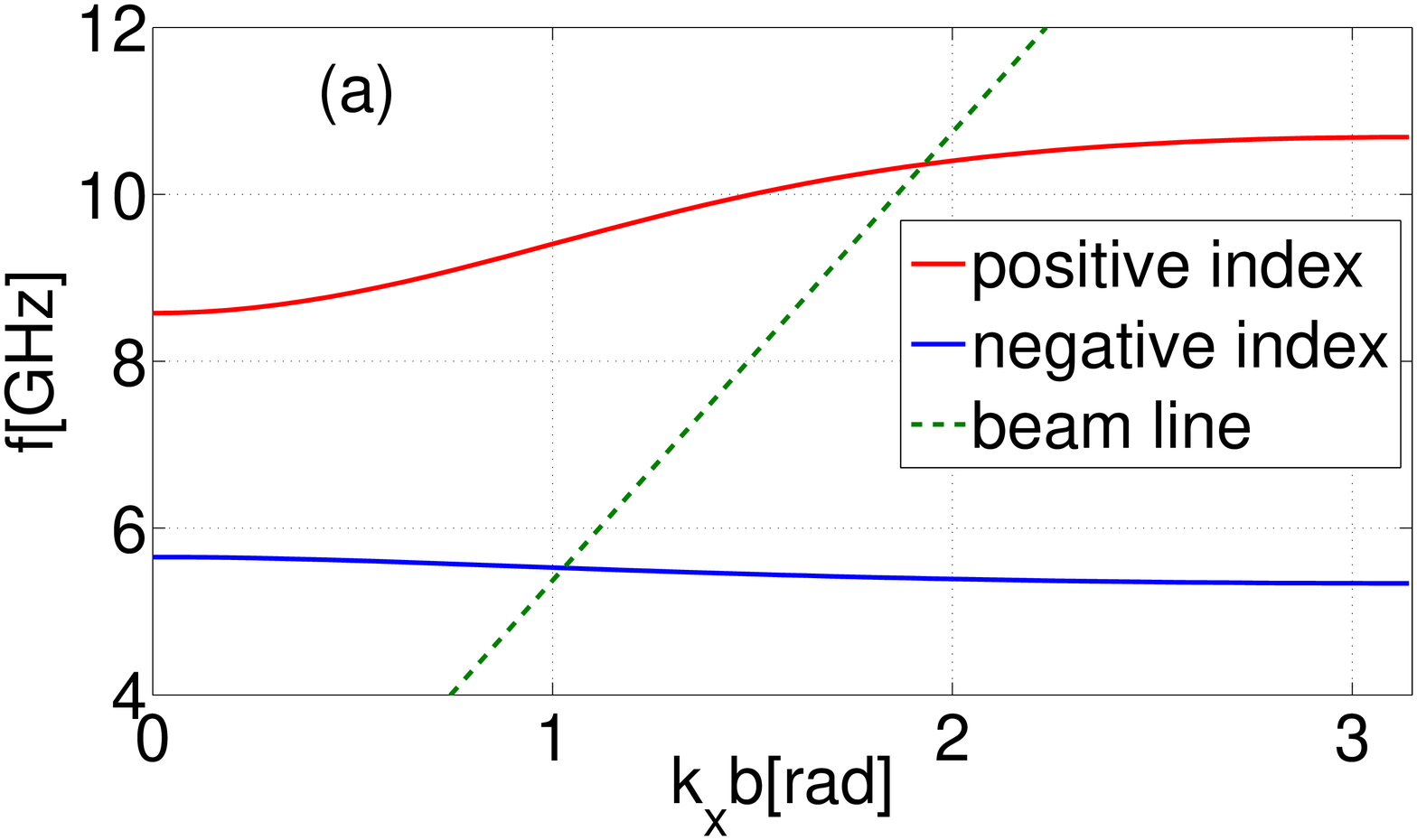}
\includegraphics[width=0.45\textwidth]{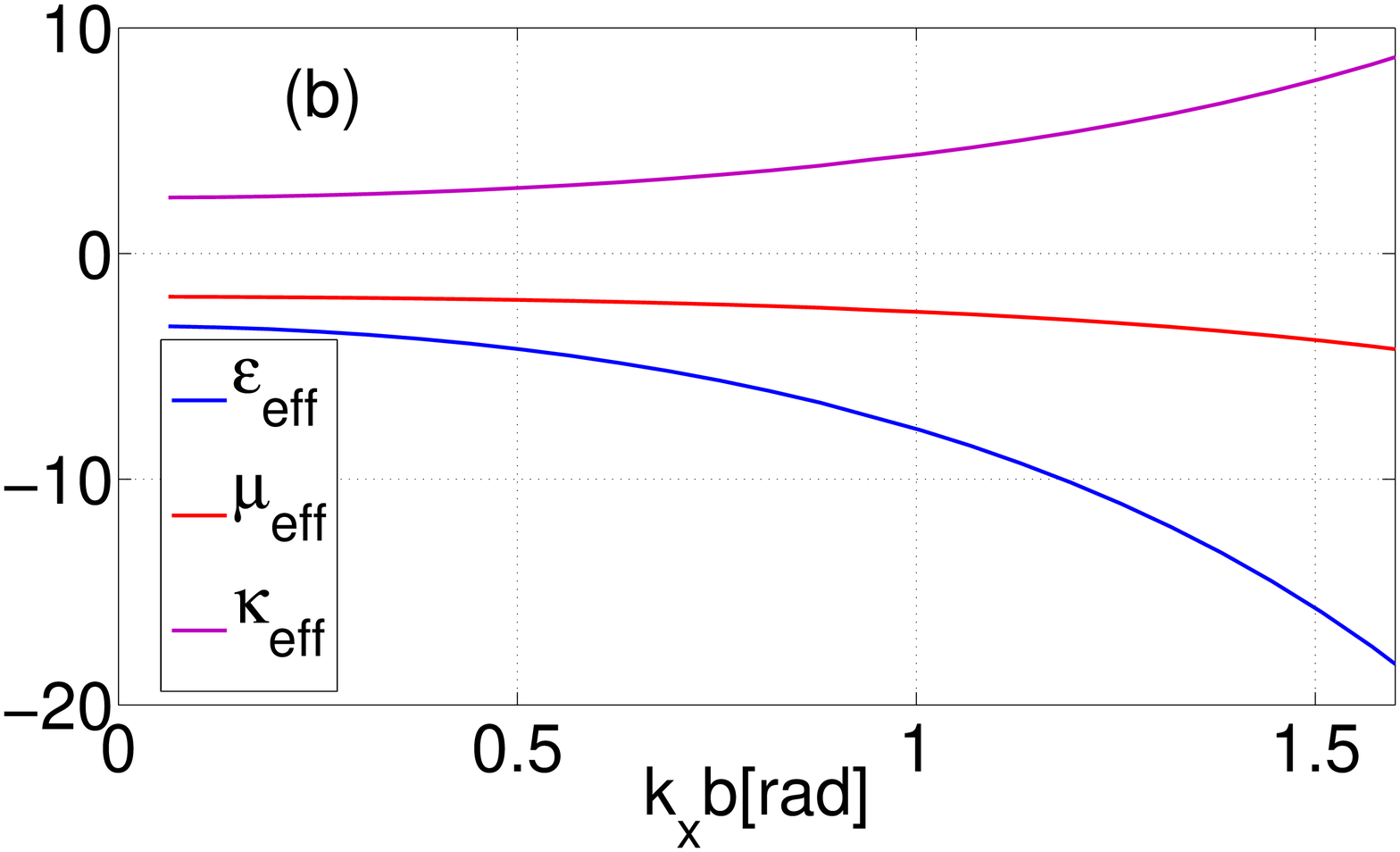}
\caption{(Color online) Modes of the NIMW calculated from COMSOL
simulations. (a) Dispersion relation for the lowest TM$_1$-like
modes (solid lines) and the "beam mode" (dashed line) defined by
$\omega = k_x v_b$, where $v_b = 0.9c$. (b) Extracted effective
parameters of the negative index mode.}\label{fig:dispersion}
\end{figure}

After gaining significant physical insights from analytic modeling of a
smooth-walled structure, we proceed to extract the constitutive
parameters of the NIMW shown in Fig.~\ref{fig:schematic} through
first-principles electromagnetic simulations using COMSOL
Multiphysics~\cite{comsol}. Periodic boundary conditions along the
$y$ and $z$ directions are used, and finite per-cell phase shift
$\Phi_x \equiv k_x b$ is assumed in the $x$-direction. While the
present design is for microwave frequencies ($f\approx f_0 = 5$
GHz; physical dimensions are given in Fig.~\ref{fig:schematic}),
it can be scaled down to mm-wave/THz frequencies. The dispersion
relations of the lowest-order modes are shown in
Fig.~\ref{fig:dispersion}(a). A narrow-band negative index (NI)
TM$_1$-like mode is indeed found in the $5.33{\rm GHz} < f <
5.65{\rm GHz}$ frequency range located below the cutoff frequency
$f_c \equiv \omega_c/2\pi \approx 11.7{\rm GHz}$. Note that a
second sub-cutoff TM$_1$-like mode with positive refractive index is also
supported by the structure. The positive index (PI) mode's
propagation is due to higher-order {\it magnetic} resonance of the
C-SRR around $11$GHz. This resonance strongly affects
$\tilde{\mu}_{yy}$ that enters Eq.~(\ref{eq:effective_params_WG})
($\tilde{\mu}_{yy} \approx 1$ is  assumed for the NI mode) and
enables $\mu_{\rm eff} > 0$ for $f > 8.5$GHz. Detailed discussion
of the PI mode is outside of the scope of this Letter, and we
concentrate below on the NI mode.

The mode-specific effective parameters of the NI mode were
extracted by applying
Eqs.~(\ref{I_and_V},\ref{eq:effective_params_TL}) to
COMSOL-produced electromagnetic field profiles and plotted in
Fig.~\ref{fig:dispersion}(b) for moderate phase advances. We note
that $\mu_{\rm eff}$ remains relatively flat, consistent with our
original conjecture that the transverse confinement of the mode is
responsible for its effective negative permeability. On the other
hand, $\epsilon_{\rm eff}$ displays strongly dispersive behavior,
consistent with its origin stemming from the resonant C-SRR
element. We further observe that the bi-anisotropy coefficient
$\kappa_{\rm eff}$ is rather large and, consistent with
Eq.~(\ref{dispersion}), explains why both $\epsilon_{\rm eff}$ and
$\mu_{\rm eff}$ are non-vanishing (negative) at the $k_x=0$
(cutoff) point, where $\epsilon_{\rm eff} \mu_{\rm eff} =
\kappa_{\rm eff}^2$ is satisfied.

\begin{figure}
\begin{tabular}{c}
\includegraphics[width=80mm]{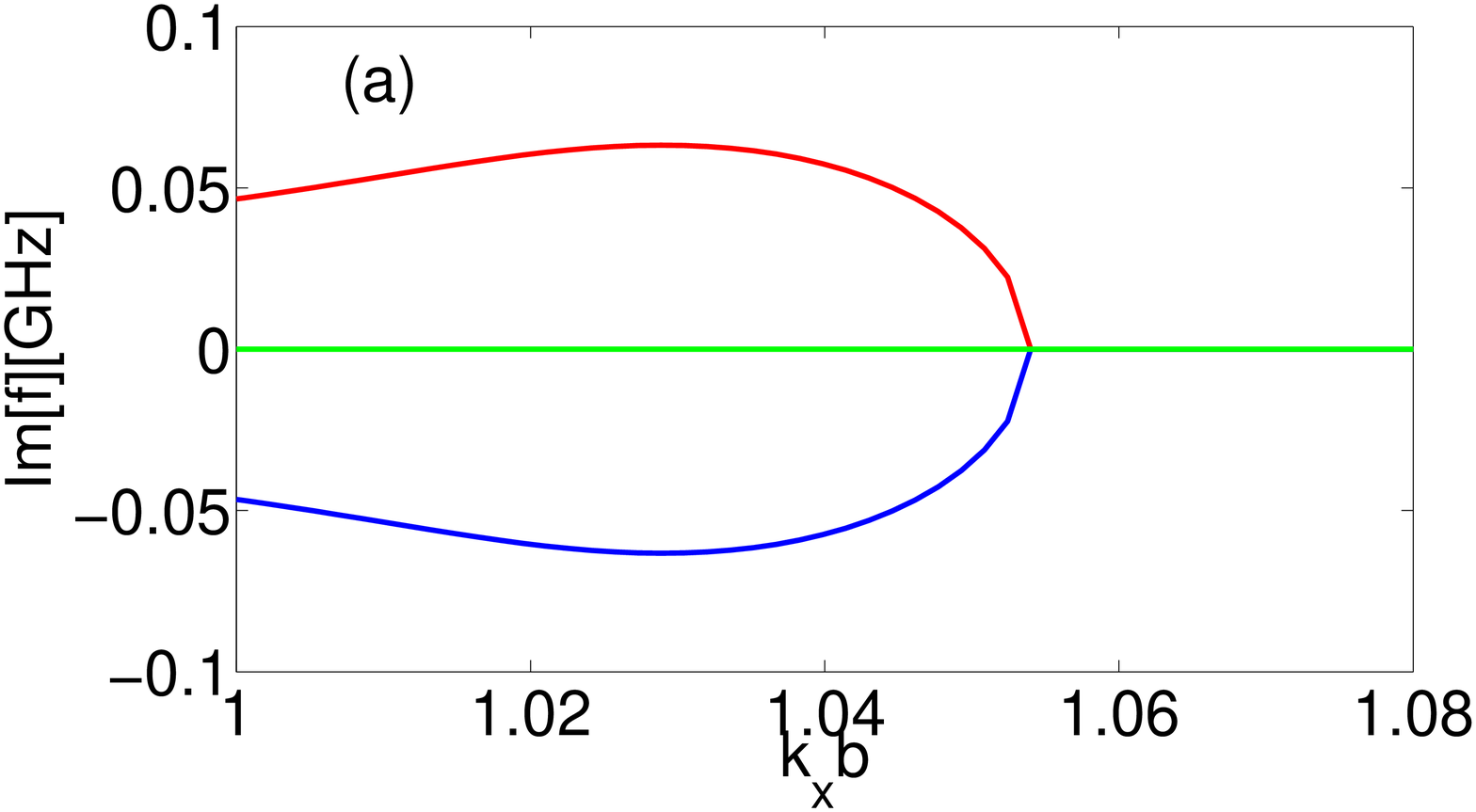} \\
\includegraphics[width=80mm]{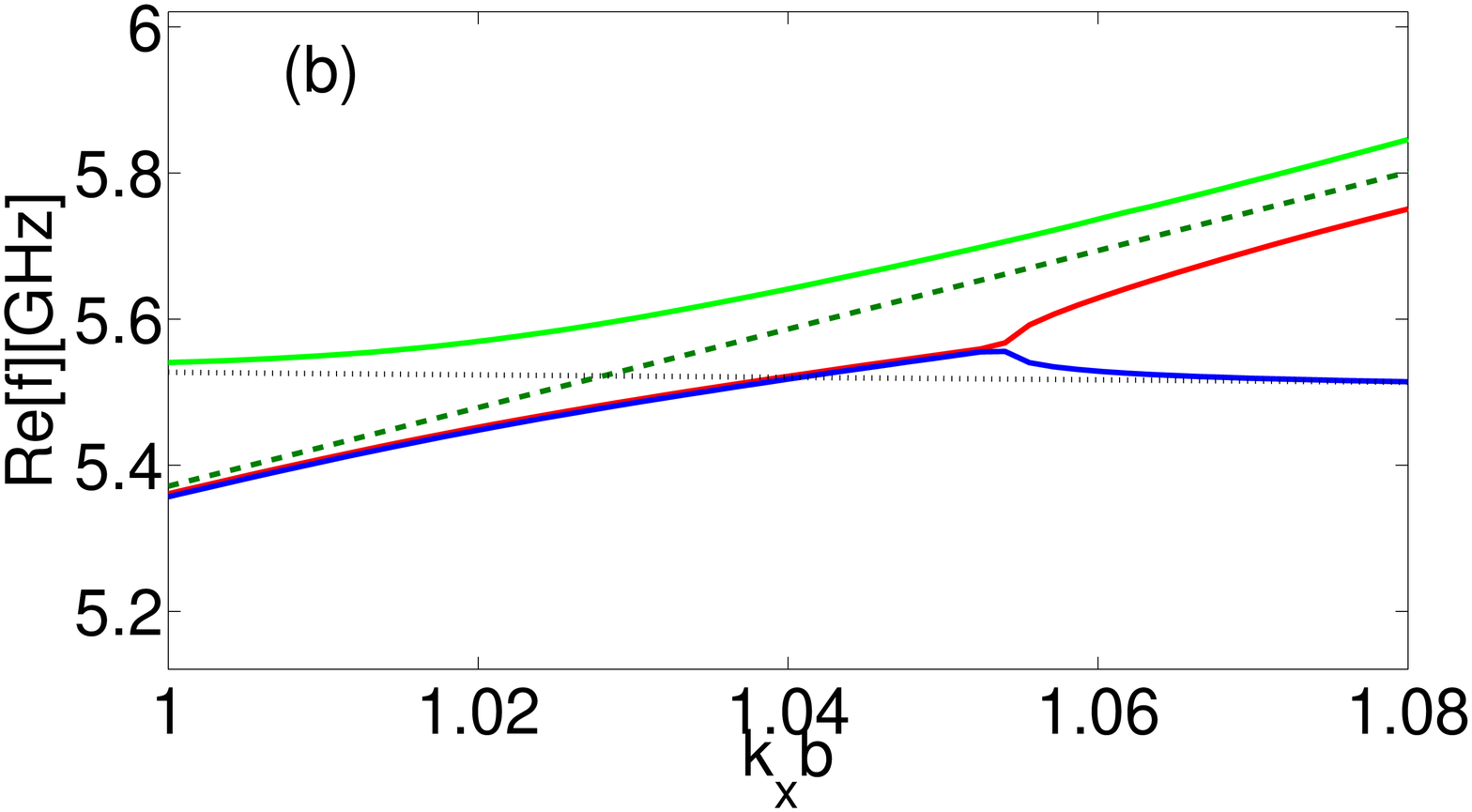}
\end{tabular}
\caption{(Color online) Dispersion characteristics of the active
negative index TM$_1$-like wave in a NIMW coupled to an electron
beam. NIMW parameters: same as in Fig.~\ref{fig:schematic}. Beam
parameters are given in the text. (a) Imaginary and (b) real parts
of the frequency as a function of the per-cell phase shift for
three hybridized modes: growing, decaying, and neutral.
Dotted/dashed lines: dispersion characteristics of the uncoupled
NIMW/beam modes. All curves are plotted in the vicinity of the
synchronous beam-mode interaction point defined by
$Re[\omega]=k_{x}v_{b}$.}\label{fig:instab}
\end{figure}

To examine the possibility of creating an active negative index
metamaterial using a high-current electron beam coupled into the
NIMW and to confirm the analytical predictions of
Eqs.~(\ref{eq:disp_eq_beam},\ref{eq:max_incr}), we have carried
out COMSOL simulations of the NIMW structure containing an
electron beam in the middle of the unit cell. The beam's presence
was modelled by assigning $\tilde\epsilon_{xx}^{(b)}$ to the
region occupied by the beam, and by assuming the following beam
parameters: $v_b=0.9c$, beam plasma frequency $\omega_b=0.01(2\pi
f_0)$, and the beam's radius $R=d/4$. The resulting complex
$\omega$, plotted as a function of the phase advance across the
cell, is shown in Fig.~\ref{fig:instab} for phase advances in the
vicinity of the beam-mode synchronism condition.

Three distinct complex $\omega$'s are found for each value of
$k_x$. Modal degeneracies can be classified according to the
value of the detuning parameter $\nu \equiv \omega_{\rm NIMW} -
k_x v_b$. For $\nu > -3\rho/\sqrt[3]{4}$ two "slow" modes with
$Re[\omega]/k_{x} < v_{b}$ degenerate in $Re[\omega]$ are found,
one of them exponentially growing and the other one decaying. The
third, "fast" mode with with $Re[\omega]/k_{x} > v_{b}$ is neutral
(neither growing nor decaying) for $\nu > 0$. For $\nu <
-3\rho/\sqrt[3]{4}$ all three modes (two "slow" and one "fast")
become neutral and non-degenerate in $Re[\omega]$. These numerical
COMSOL results compare very well with the analytical predictions
of Eq.~(\ref{eq:disp_eq_beam}) obtained by adjusting the effective
beam plasma frequency to $\omega_b^{eff}=0.05 \omega_b$ to account
for only partial overlap between the beam and the negative-index
TM mode. This reduction in $\omega_b^{eff}$ is associated with
small shunt impedance of the resonant NIMW, which concentrates the
electric energy away from the beam in the vicinity of the C-SSR.

In conclusion, we have demonstrated a geometry to realize a novel
active beam-driven negative index meta-waveguide (NIMW) that
supports transverse magnetic (TM) waves capable of resonantly
interacting with an electron beam. A number of novel vacuum
electronics devices that require backward waves and small group
velocity, such as backward-wave oscillators and amplifiers, can be
envisioned based on this concept. The sub-wavelength nature of the
unit cell enables strong interaction with electron beams at the
fundamental harmonic of the structure, while the resonant nature
of the constitutive elements (complementary split ring resonators)
enables low group velocity and, potentially, agile frequency
tuning. The narrow bandwith and small group velocity of NIMW
increases its shunt impedance, making it a potentially attractive
structure for advanced accelerator application. This work is
supported by the US DoE grants DE-FG02-04ER41321 and
DE-FG02-91ER40648.

\end{document}